\documentclass[letterpaper, 10 pt, peerreviewca]{ieeeconf}

\IEEEoverridecommandlockouts
\usepackage{graphicx,subfigure}
\usepackage{epstopdf}
\usepackage{tikz}
\usepackage{amsmath}
\usepackage{multirow}
\usepackage{amsfonts}
\usepackage{float}
\usepackage{amssymb}
\usepackage{graphicx}
\newtheorem{theorem}{Theorem}

\newtheorem{proposition}{Proposition}

\makeatletter

\newcommand{\Rmnum}[1]{\expandafter\@slowromancap\romannumeral #1@}
\makeatother

\begin{document}
\title{\LARGE  Paradigm and Paradox in Topology Control of Power Grids}
\author{Shuai Wang \& John Baillieul}
\maketitle
\let\thefootnote\relax\footnotetext{\noindent\underbar{\hspace{0.8in}}\\
Shuai Wang is  with the Division of Systems Engineering and John Baillieul is with the Departments of Mechanical Engineering; Electrical and Computer Engineering, and the Division of Systems Engineering at Boston University, Boston, MA 02115. Corresponding author is John Baillieul (Email: johnb@bu.edu). \newline The authors gratefully acknowledge support of NSF EFRI grant number 1038230. }
\begin{abstract}  Corrective Transmission Switching  can be used by the grid operator to relieve line overloading and voltage violations, improve system reliability, and reduce system losses.
Power grid optimization by means of line switching is typically formulated as a mixed integer programming problem (MIP).  Such problems are known to be computationally intractable, and accordingly, a number of heuristic approaches to grid topology reconfiguration have been proposed in the power systems literature. By means of some low order examples (3-bus systems), it is shown that within a reasonably large class of ``greedy'' heuristics, none can be found that perform better than the others across all grid topologies. Despite this cautionary tale, statistical evidence based on a large number of simulations using using IEEE 118-
bus systems indicates that among three heuristics, a globally greedy heuristic is the most computationally
intensive, but has the best chance of reducing generation costs while enforcing N-1 connectivity. It is argued that, among all iterative methods, the locally optimal switches at each stage  have a better chance in not only approximating a global optimal solution but also greatly limiting the number of lines that are switched.
\end{abstract}
 \section{Introduction}
Power grid control by means of transmission
line switching has been the focus of research
since the 1980s \cite{Koglin}, \cite{Van} in the hope of taking advantage of its fast time constants in changing the system state to reduce losses and improve grid security. After the adoption  of energy deregulation in   the 1990s, a good deal of effort in the study of power markets has been focused on the co-optimization of generation unit commitment and transmission line
switching when  congestion occurs. 

A standard and widely adopted approach to modeling topology reconfiguration  uses   binary variables to indicate the status of candidate lines (in or out of service) in a mixed integer programming (MIP) formulation   \cite{Chen}. To address the complexity caused by the  NP-hardness of MIP, most of recent work  focuses on developing  fast heuristic algorithms. References  \cite{Hedman,Fisher, Ruiz, Fuller, Soroush} show the effectiveness of heuristics co-optimizing the generation and the network topology through simulations on the IEEE 118-bus system and the the WECC 179-bus system, and the  running time records of such heuristics have been consistently set and then shattered.  At the same time, the quest for
firmer theoretical foundations of the heuristic approaches continues.  References \cite{Baillieul} and \cite{Wang}, for instance, show that in a voltage-controlled or current-controlled  DC circuit, the sign of the overall $I^2R$ congestion change is always predictable. Unfortunately, the magnitude of change can only be determined by solving Kirchhoff equations.
 
Some NP-complete or NP-hard problems  may admit efficient heuristics in practice as long as their inputs meet certain criteria. For example, there is a pseudo-polynomial time algorithm using dynamic programming for the {\em knapsack problem} if the inputs are of relatively small magnitude \cite{Kellerer}, and the {\em Fincke and Pohst algorithm} \cite{Fincke} is suitable for solving integer programming with pseudo continuous inputs, etc. There are, however, no such  algorithms   for general MIP problems, indicating a strong form of NP-hardness for MIP. 

The present paper examines fundamental challenges associated with heuristics for topology reconfiguration of power grids.  The paper is organized as follows. We begin the discussion in the next section by explicitly showing the NP-hardness of topology control with a simple proof.  In Section III, we explore a list of paradoxical behaviors associated with greedy heuristics for topology control, and describe examples showing that many plausible heuristic approaches fail to achieve acceptable grid operation. In particular,  we define the concepts of $commutativity$ which is the ease of creating an optimal sequence of operations out of optimal sub-sequences, and   $monotonicity$ which is the possibility of continued step-wise improvement on the effects of previous operations, and $consistency$ which is the ease of obtaining better operation by increasing the computational complexity.  Simulation results in Section IV indicate that the shortcomings of heuristics
pointed out in Section III are statistically insignificant across a wide range
of power grids that are frequently studied in the power systems literature. The section ends with a theoretical justification in supporting the value of greedy heuristic in topology control. Concluding remarks   are contained in Section V. 
\section{The NP-hardness of Topology Control}
Mathematically, the DC model of  power flow is  equivalent to a current driven network, where power injections are equivalent to  current sources; power flowing through lines is equivalent to current through edges, etc.  See Table 1. 

Reference \cite{Wang}  approaches the topology control problem by combining graph theory and Kirchhoff laws. The work in \cite{Wang} shows that disconnecting a resistance edge in a current driven network always increases the total $I^2R$ congestion of the network. In other words, the new circuit network obtained after line switching has fewer edges but has higher overall $I^2R$ congestion.

\begin{center}
\begin{tabular}{l@{\hskip 0.1in}c@{\hskip 0.1in}c@{\hskip 0.1in}c@{\hskip 0.1in}c}\hline
 network &  potential   &  injection  &  admittance  & equation \\ \hline 
 circuit & voltage $V$ & current $I$ & conductance $G$ & $I=GV$   \\  \hline
grid&  phase $\theta$  & power $P$ & susceptance $B$ &   $P=B\theta$ \\  \hline
\end{tabular}\end{center}\begin{center}
{Table 1: The equivalence between a current driven circuit and a transmission grid.}
\end{center}

By the   equivalence  between  the  current-controlled  circuits  and  DC  power  flow  models  of transmission grids,  we know the total real $ f^2/B$ (squared line flow over line susceptance) of the transmission network must be increased after switching off a transmission line. Defining   $\overline{f^2}/B$   as the squared line flow limit over line susceptance, we find that the total $\overline{f^2}/B$ capacity must strictly decrease after disconnecting a line. This means that the total $f^2/B$ stability margin, the difference between the total  $\overline{f^2}/B$ capacities and actual real  $f^2/B$, must be decreased after each step switching out lines.   This also suggests a good rule of thumb of line switching in transmission networks: do not switch out lines in a congested network with relatively low  $\overline{f^2}/B$ stability margin on those uncongested lines.
 But what if we are dealing with a congested network with high capacity margin on uncongested lines?

Topology control is typically formulated as a mixed integer  programming (MIP) problem \cite{Chen}.  Although the common wisdom is that many MIP problems are NP-hard and some of them don't even have  pseudo-polynomial time solutions, an explicit proof of the NP-hardness of topology control, to the best of our knowledge, is somehow lacking in the literature. Thus, we start the  discussion by answering this question by proving the following.
\begin{proposition}
Given the capacity of each line and the net power injection/withdrawal of each bus in an arbitrary DC model of a  power network, the  topology control problem is NP-complete. \label{nphard}
\end{proposition}

\textit{Proof:}
First we know that the feasibility problem of topology control is NP because we can verify in polynomial time whether an instance is a feasible solution. The verification involves two steps. The first is to calculte the new line flows  by solving the updated OPF problem. Next, we check whether line congestions  still exist. Clearly, both steps can be done in polynomial time.

The second part of the  proof involves a reduction from an arbitrary instance of an NP complete problem to an instance of the topology control problem. 

Here, we use the well known \emph{subset sum problem} for the reduction. The subset sub problem is this: given a set of numbers, is there a non-empty subset whose sum is zero? For instance, given the set $\{-1, -2, -3, 4, 8\}$, the answer is yes because the subset $\{-1, -3, 4\}$ sums to zero. The problem is known to be NP-complete \cite{Murty}. We   take an instance of the subset sum problem and reduce it to a topology control instance that has a feasible solution if and only if the subset sum problem has a non-empty subset whose sum is zero.

\begin{figure}[htbp]
	\centering
		\includegraphics[width=0.35\textwidth]{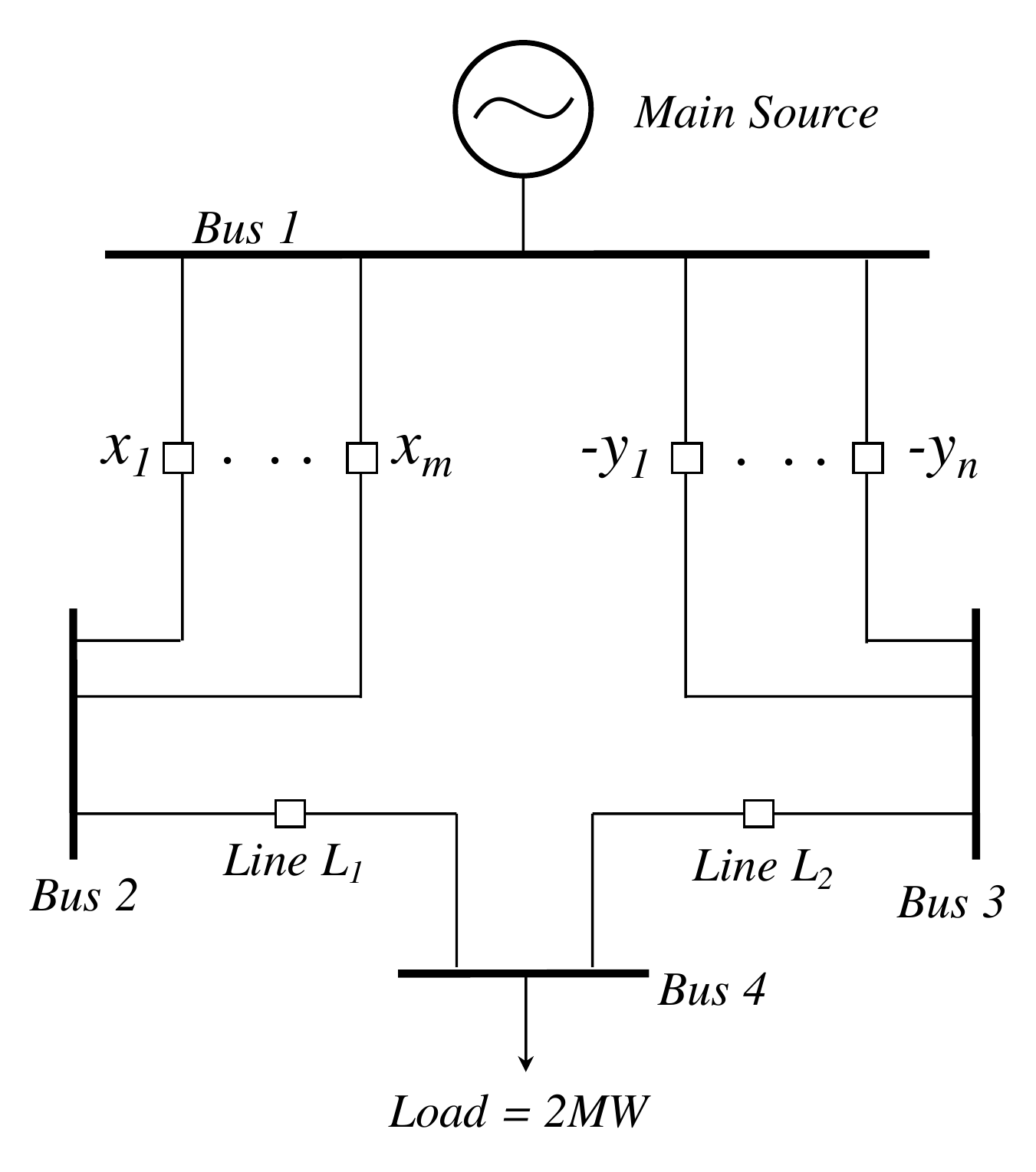}
	\caption{An instance of topology control problem.} \label{fig:np-hard}
\end{figure}

Let $X=\{x_1,...,x_m,y_1,...,y_n\}$ ($x_i>0$, $i=1,...,m$ and $y_j<0$, $j=1,...,n$) be an instance of the subset sum problem. We then can reduce it to the topology control problem shown in Fig. \ref{fig:np-hard}. 

In Fig. \ref{fig:np-hard}, we assume that the susceptances of vertical lines are $\{x_1,...,x_m\}$ ($x_i>0$, $i=1,...,m$) and $\{-y_1,...,-y_n\}$ ($y_j<0$, $j=1,...,n$), respectively. All other lines are assumed to be of same susceptance. In addition, the line capacities of the lowest two lines $\{L_1,L_2\}$  are both 1 $MW$, and we assume the line capacities of other lines are all large enough. To satisfy the 2 $MW$ demand of the lower bus, we must balance the susceptance of the left group of lines $\{x_1,...,x_m\}$  with that of the right group of lines $ \{-y_1,...,-y_n\} $ . Then we see that the topology control instance has a feasible solution if and only if the subset sum problem $\{x_1,...,x_m,y_1,...,y_n\}$ has a non-empty subset whose sum is zero.$\hfill{} \Box$

\section{Paradox of Topology Control}\label{sec:paradox}
Proposition \ref{nphard} is not surprising as most  integer programming is NP-hard. The NP-hardness naturally leads researchers to think about possible heuristic methods. In recent years there has been a significant interest in developing heuristics to co-optimize network topology and generation  \cite{Fisher,Ruiz,Fuller,Soroush}. For example, the work reported in \cite{Ruiz}  finds topology improvements by iteratively solving the linear programming (LP) formulation of the DC optimal power flow (DCOPF) by means of a heuristic algorithm that disconnects the one
unprofitable  line   per iteration while enforcing a relaxed form of  N-1 reliability.
Such a heuristic approach is one way of avoiding the severe
computational complexity of the
line switching problems, and makes the real time topology control   possible. The NP-hardness of the problem, however, makes optimality not guaranteed by any heuristic, and we are led to a set of interesting paradoxical phenomena.

We  focus our attention on several iterative greedy heuristics in which the main idea is to disconnect/connect the line(s) that can reduce the cost most in each iteration. We show  paradoxical behaviors associated with the heuristics by using some simple 3-bus power system examples.

\subsection{Non-commutativity}
The first paradoxical behavior is  that the most beneficial single line may not be included in the most beneficial pair of lines or the most beneficial set of $n$ ($n>2$) lines. We call this    the $non$-$commutativity$ property of the topology control problem. This problem can be easily illustrated as follows. 
\begin{figure}[htbp]
	\centering
		\includegraphics[width=0.5\textwidth]{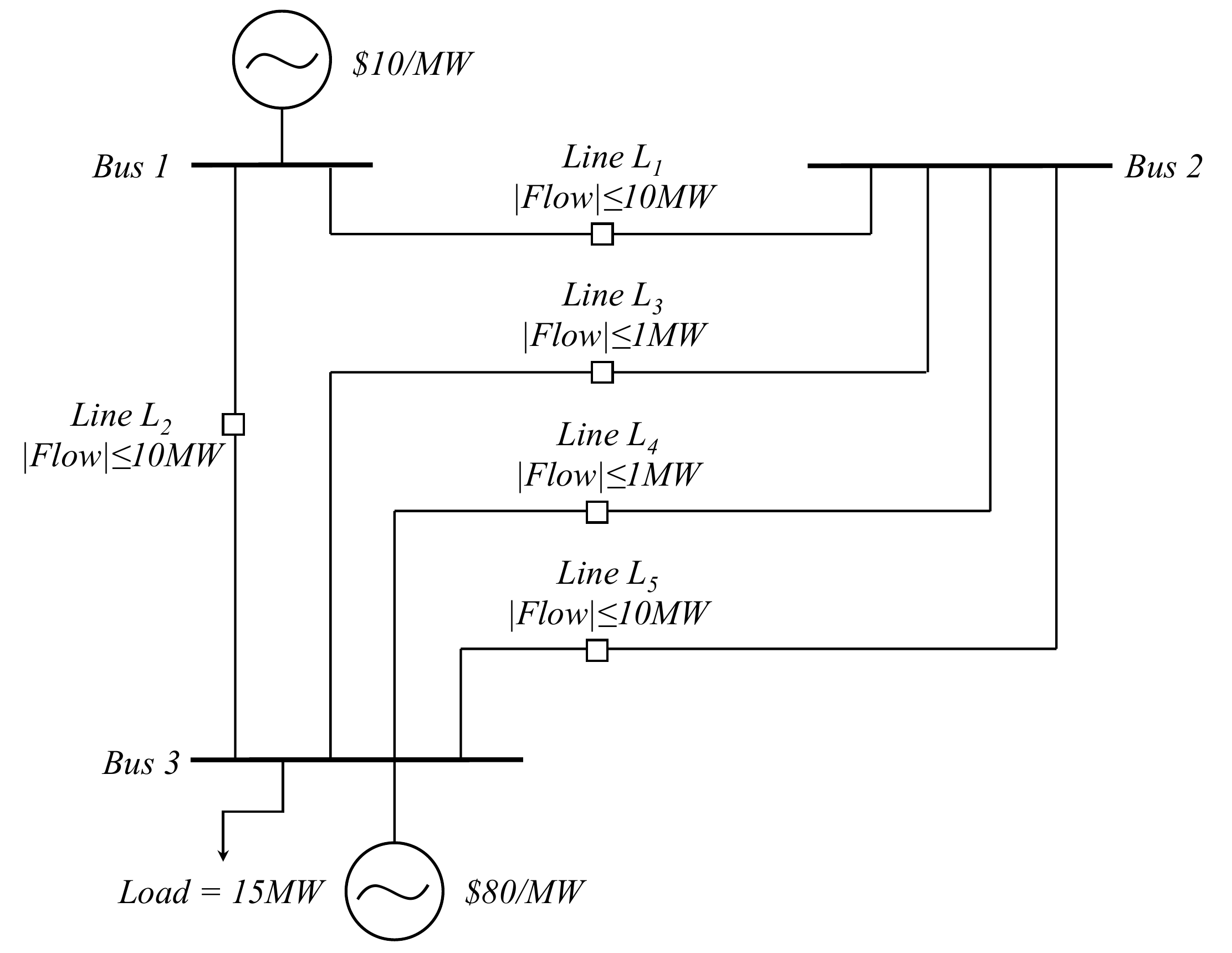}
	\caption{The example of non-commutativity problem.} \label{fig:non-commutativity}
\end{figure}

In Fig. \ref{fig:non-commutativity}, there are two generators with the cheap one on   bus 1 and the  expensive one on bus 3 and a 15MW demand on bus 3.  We assume that all transmission lines have the same susceptance, and the line capacities range from 1MW to 10MW as marked in the figure. The original cost before applying topology control is $\$$710 with the cheap generator supplying 7MW and the expensive generator supplying 8MW.

If we  remove the most  beneficial single line per iteration, then we will remove the line L1  and the cost will be $\$$500  with the cheap generator supplying 10MW and the expensive generator supplying 5MW. The algorithm will stop here as we can't further reduce the cost by removing another line. However, if we are allowed to remove the most  beneficial pair of lines per iteration, then we will keep L1 connected and instead remove the line L3 and L4 together  and the cost will be $\$$150  which actually equals  the cost of  unconstrained optimal power flow. 
\subsection{Non-monotonicity}
The second paradoxical behavior is that  one line which was removed earlier may be reconnected later because keeping it switched off is no longer beneficial. We call this   the $non$-$monotonicity$ property of the topology control problem. 
\begin{figure}[htbp]
	\centering
		\includegraphics[width=0.55 \textwidth]{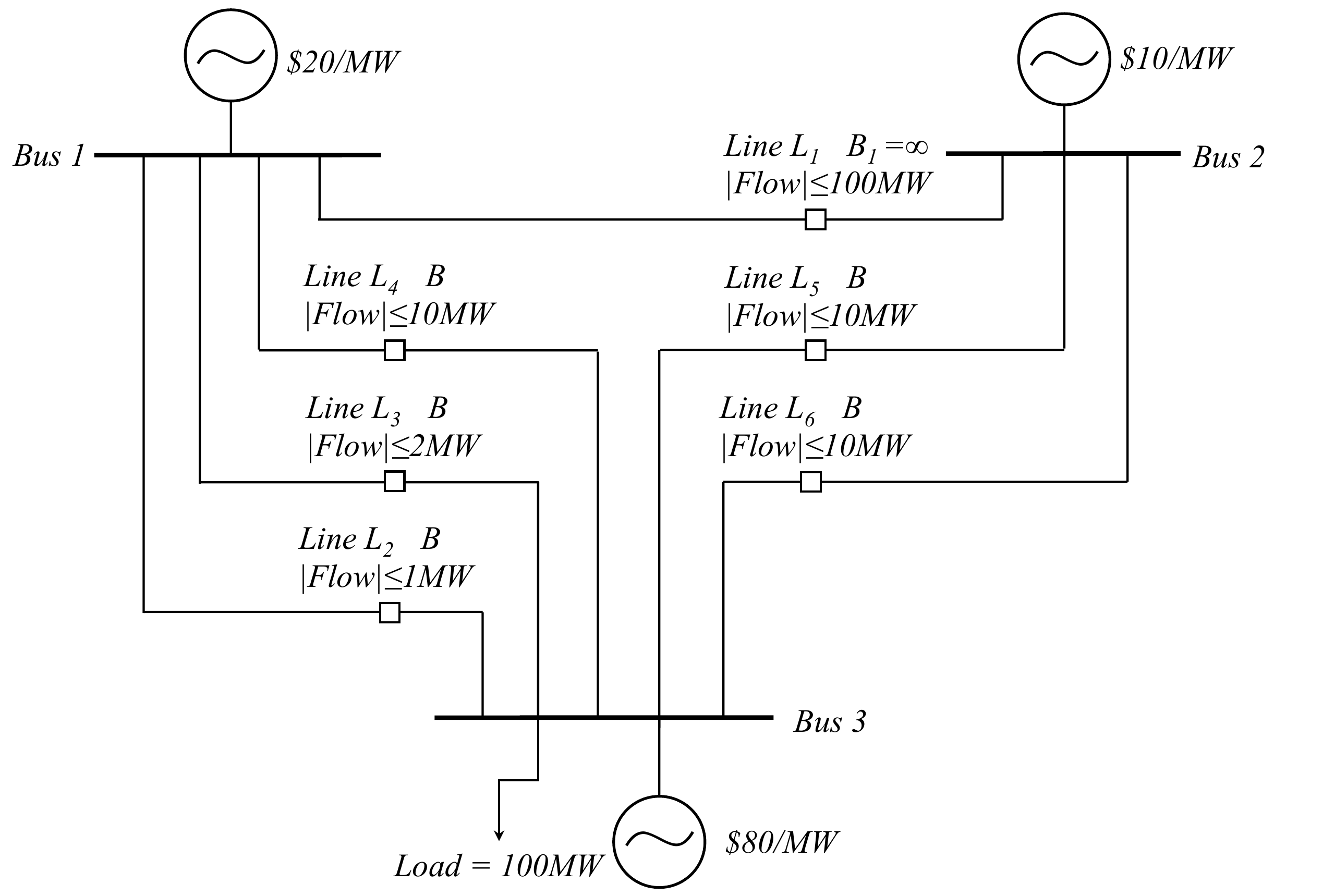}
	\caption{The example of non-monotonicity.} \label{fig:non-monotonicity}
\end{figure}

This problem is illustrated in Fig. \ref{fig:non-monotonicity} which shows a power system with the cheapest generator on bus 2, a slightly more expensive  generator on bus 1, and the most expensive generator on bus 3. We assume that the susceptance of the horizontal transmission line is large enough and all other lines are of the same but limited susceptance. There is a demand of 100MW on bus 3 and the line capacities range from 1MW to 100MW as marked in the figure. The original cost before applying topology control is $\$$7650 with the cheapest generator supplying 5MW and the most expensive generator supplying 95MW.

Suppose we can either remove or connect the most beneficial  single line per iteration, then we will  remove the line L1 first, line L2 second, and line L3 third. In the fourth iteration, however, we find that having L1 out of service is no longer beneficial and we can reduce the cost by reconnecting L1. After the reconnecting, the  cost  would be $\$$5900 with the cheapest generator supplying 30MW  and the most expensive generator supplying 70MW. The algorithm will stop here as we can't further reduce the cost by removing or reconnecting another line.
\subsection{Non-consistency}
\begin{figure}[htbp]
	\centering
		\includegraphics[width=0.55 \textwidth]{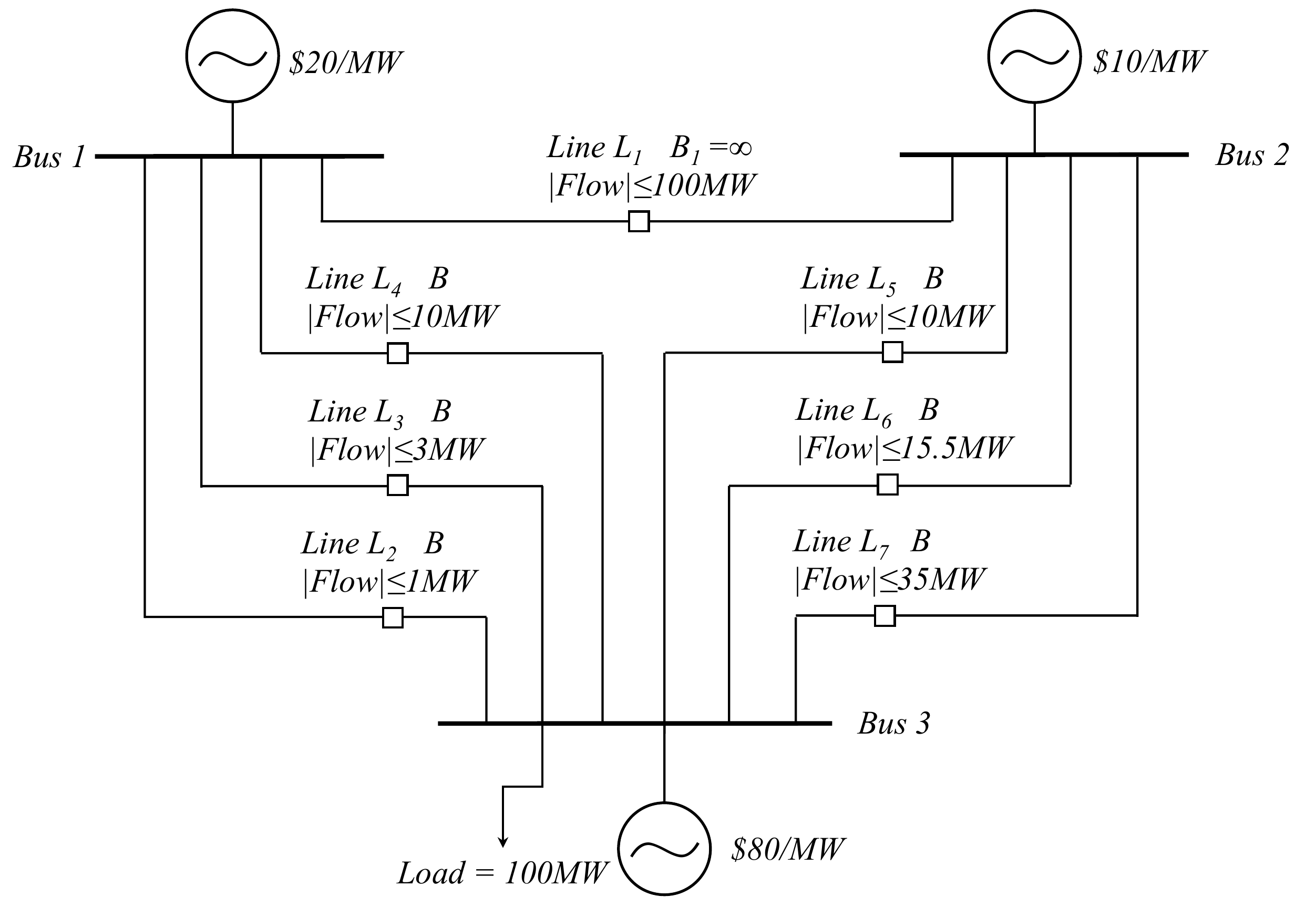}
	\caption{The first example of non-consistency problem.} \label{fig:reconnect}
\end{figure}

The third example illustrated in in Fig. \ref{fig:reconnect} shows that an  algorithm that only allows removal of one line per iteration may outperform another algorithm that allows to either remove or reconnect one line per iteration.  We call this the $non$-$consistency$ property of the topology control problem. This shows that the increased computational complexity sometimes is not compensated by lower generation cost.

Fig. \ref{fig:reconnect} shows a power system with the cheapest generator on bus 2, and the  mid-cost generator on bus 1, and the most expensive generator on bus 3. We assume that the susceptance of the horizontal transmission line is large enough and all other lines are of the same but limited susceptance. There is a demand of 100MW on bus 3 and the line capacities range from 1MW to 100MW as marked in the figure. The OPF with all lines in service is $\$$7580 with the cheapest generator supplying 6MW and the most expensive generator supplying 94MW.

Algorithm (1) (only allowing removal of one line per iteration) will  remove line L1 first, line L2 second, line L3 third,  line L5 fourth, and line L6 last. The final cost  will be $\$$4950 and with the cheapest generator supplying 35MW, and the mid-cost generator supplying 10MW, and the most expensive generator supplying 55MW. 

Algorithm (2) (allowing to either remove or reconnect one line per iteration)  will remove line L1 first, line L2 second, and line L3 third. In the fourth iteration, line L1 will be reconnected and the algorithm stops. The cost  will be $\$$5200 with the cheapest generator supplying 40MW and  the most expensive generator supplying 60MW. 

As we can see algorithm (1) outperforms algorithm (2).\\

\begin{figure}[htbp]
	\centering
		\includegraphics[width=0.45 \textwidth]{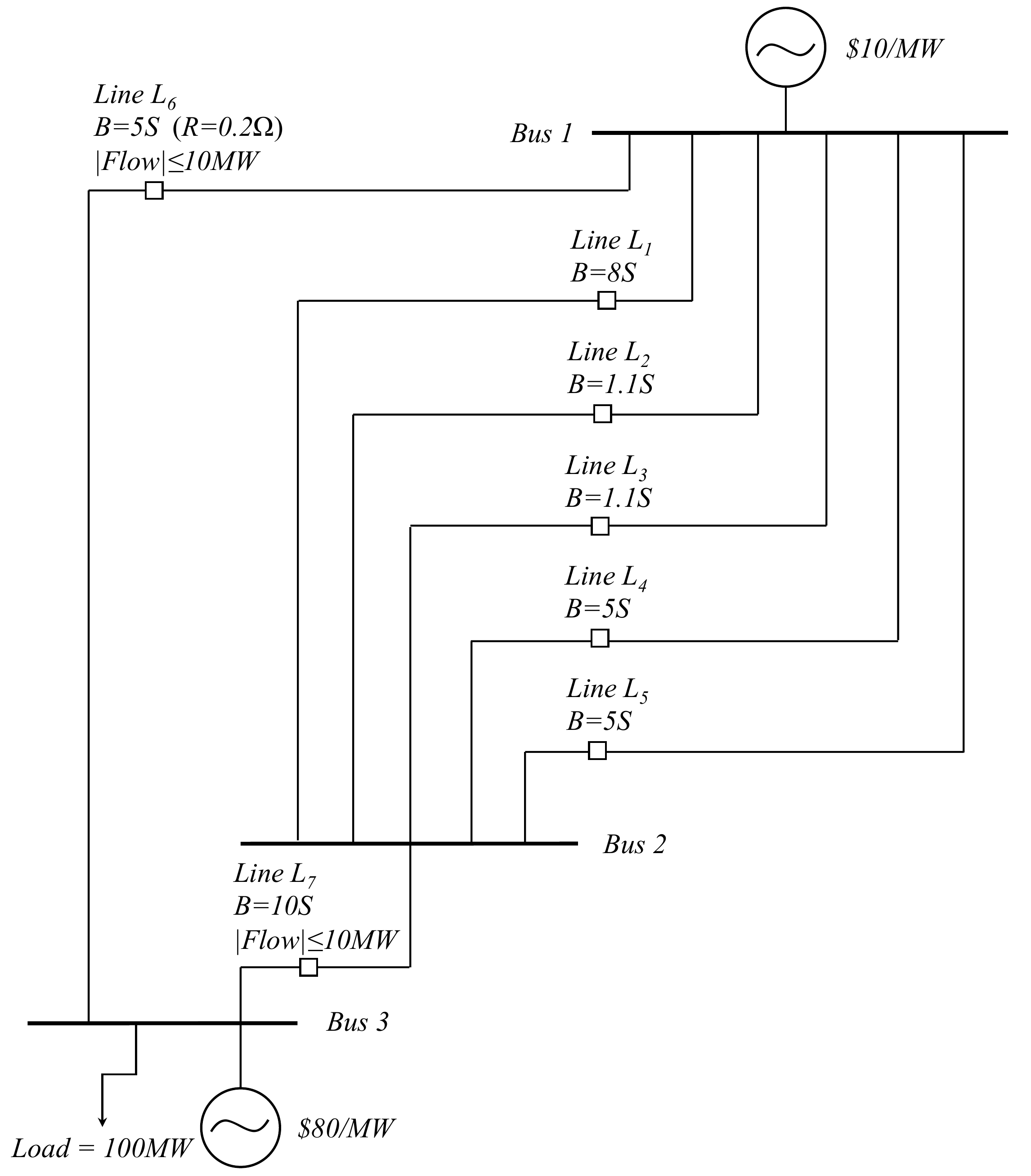}
	\caption{The second example  of non-consistency problem.} \label{fig:increasenumber}
\end{figure}

The fourth example illustrated in Fig. \ref{fig:increasenumber} shows another type of non-consistency property: one algorithm  that only allows removal of one line per iteration may outperform another algorithm that allows removal of at most 2 lines line per iteration. 

Fig. \ref{fig:increasenumber} shows a power system with the cheap generator on the upper bus and the  expensive generator on the lower bus. We assume that the susceptances of L1 through L7 are 8S, 1.1S, 1.1S, 5S, 5S, 5S and 10S, respectively. There is a demand of 100MW on the lower bus and the line capacities of L6 and L7 are both 10MW. The line capacities of other lines are assumed to be sufficiently large. The OPF with all lines in service is $\$$6775 with the cheap generator supplying 17.5MW and the expensive generator supplying 82.5MW.

Algorithm (1) (only allowing removal of one line per iteration) will remove line L1 first, line L2 second, and line L3 third. The cost  will be $\$$6600 with the cheap generator supplying 20MW and the  expensive generator supplying 80MW. The algorithm will stop here as we can't further reduce the cost by removing another line.

Algorithm (3) (allowing to remove at most two lines per iteration) will remove line L4 and L5 in the first iteration, with the resulting cost being $\$$6607 with the cheap generator supplying 19.9MW and the  expensive generator supplying 80.1MW. The algorithm will stop here as we can't further reduce the cost by removing  another line or another pair of lines.

As we can see algorithm (1) outperforms algorithm (3). 
\\

\begin{figure}[htbp]
	\centering
		\includegraphics[width=0.4\textwidth]{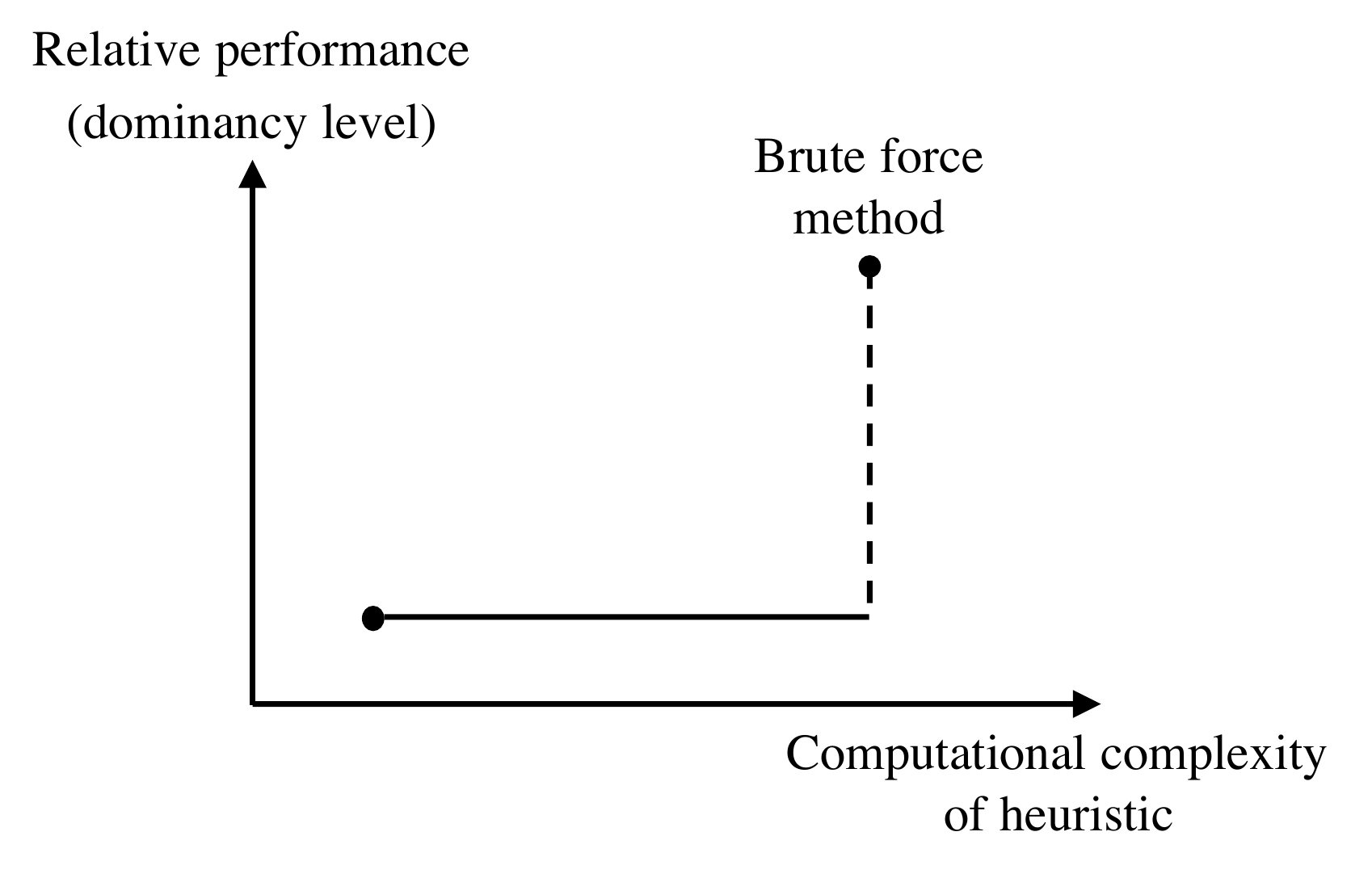}
	\caption{Visualization of the paradoxical phenomenon: no extra return for increased computational complexity of heuristics in topology control.} \label{paradox}
\end{figure}

Mathematically,  the Algorithms (1), (2) and (3) described above are of the same level of performance, i.e. no one  dominates another. For a general binary programming problem,  only when the search  space of one heuristic explicitly contains that of the other   can it declare dominancy. Thus the paradoxical phenomenon, {\em ``no `guaranteed' extra return for increased computational complexity"}, can be observed in all types of non-brute-force greedy heuristics. This phenomenon, which makes the binary programming problem extremely challenging, is visualized in Fig. \ref{paradox}. 

\section{Paradigm of Topology Control}\label{sec:paradigm}
Section \ref{sec:paradox} explicitly discounts the value of increased  computational effort of heuristics in reducing the generation cost for   {\em specific} grid scenarios. It is natural to ask if the next best thing is true: can the increased  computational effort  ``statistically" reduce   generation costs for a {\em large set} of scenarios?  The result of our simulations  suggests the answer is yes.

\subsection{Test Network Overview}
The heuristics were tested on the IEEE 118-bus test system. This test system represents a portion of the Midwestern US  Power System  as of December, 1962. The version of the test system employed is available at \cite{118}. The generator cost information used in the IEEE 118 network is extracted from  \cite{Blumsack}. The test system consists of 118 buses, 54 generators, and 194 lines, all of
which are assumed to be connected in the initial topology.

\subsection{Heuristics Overview}
Three heuristics were tested and compared in this section. All of them sequentially disconnect transmission lines until no further generation costs can be reduced.

To make this a fair comparison, all heuristics only open closed switches and do  not close open switches since the Line Profits Heuristic \cite{Ruiz} (described below) cannot compute potential improvements from reconnecting disconnected lines.

\subsubsection{Random Heuristic}
\begin{figure}[htbp]
	\centering
		\includegraphics[width=0.55\textwidth]{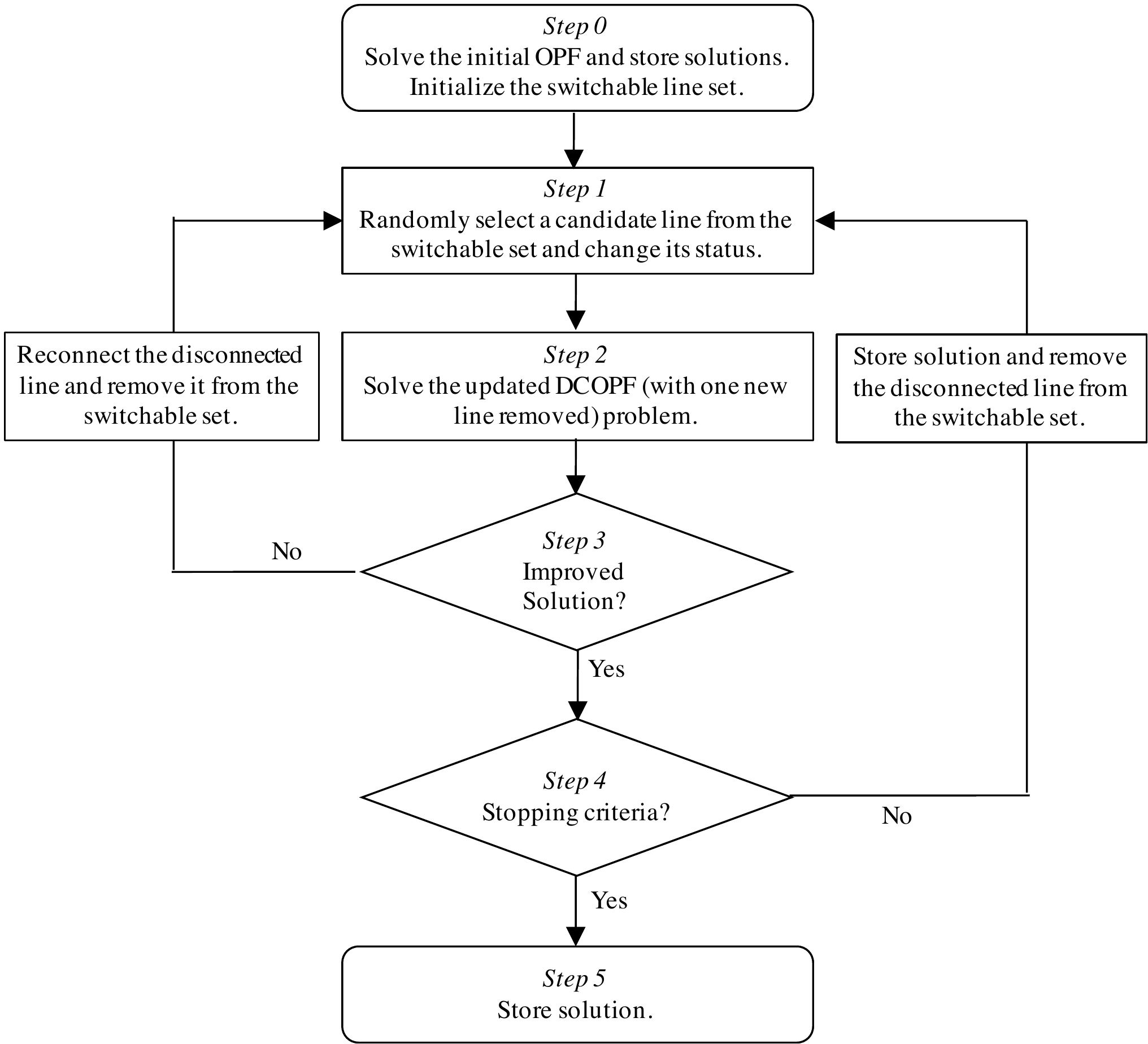}
	\caption{Flow chart describing general algorithm structure of the Random Heuristic.} \label{random}
\end{figure}

The algorithm is specified in Fig. \ref{random}. This heuristic sequentially solves the DCOPFs. In each iteration, if the DCOPF is feasible, a set of switchable candidate lines is then created.   The cardinality of the switchable set is reduced in each iteration after randomly disconnecting a switchable unprofitable line. In addition,  the switchable set enforces a relaxed form of the ``N-1" reliability requirement: at no time is a load or generator bus   connected by less than TWO lines.  The heuristic stops when no switchable unprofitable lines can be found.

\subsubsection{Line Profit Heuristic} This heuristic is proposed in \cite{Ruiz}, and is named after the metric used in the switching criteria.  Its algorithm structure is almost the same  as Fig. \ref{random} except for the line switching step (Step 1 in Fig. \ref{random}). In the line switching step, the line profits, $f_l(\pi_{n_l}-\pi_{m_l})$, are computed where $f_l$ is the  power flow on transmission line $l$, and  $\pi_{n_l}-\pi_{m_l}$ are the locational marginal price (LMP) difference between the two ending buses of line $l$. The convention is that the flow direction of line $l$ is from bus $m$ and to bus $n$. The most unprofitable line (with lowest line profit $f_l(\pi_{n_l}-\pi_{m_l})$), rather than a random line, is selected  as a candidate for opening. Please refer \cite{Ruiz} for more details.  

\subsubsection{Standard Greedy Heuristic}
\begin{figure}[htbp]
	\centering
		\includegraphics[width=0.55\textwidth]{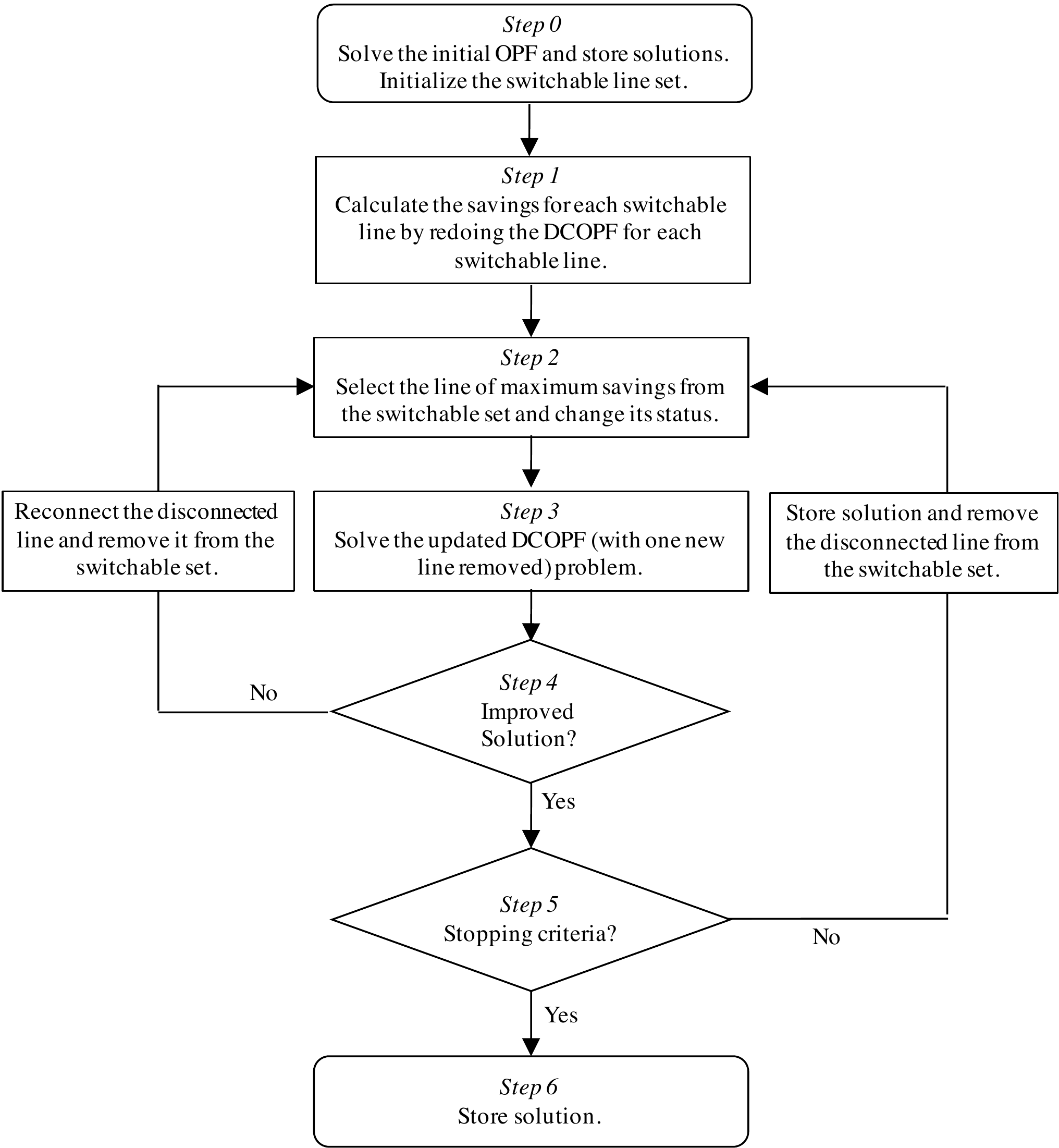}
	\caption{Flow chart describing general algorithm structure of the Random Heuristic.} \label{standard}
\end{figure}

The algorithm is specified in Fig. \ref{standard}. Unlike the Line Profit Heuristic that first sorts the candidate lines based on their  line profits $f_l(\pi_{n_l}-\pi_{m_l})$ and then selects the first line whose removal leads to  savings  in the updated DCOPF, the Standard Greedy Heuristic redoes  the updated DCOPF for each  allowable line removal and then selects the single line  to be removed that leads to maximum savings in each iteration. 

Clearly, optimality is not guaranteed with any of the heuristics, since the transmission topology  is not co-optimized with power dispatch. 
 
\subsection{Simulation Results and Analysis} \label{simlu}
To generate the test scenarios for the heuristics, a fixed load is maintained and we perform a Monte Carlo simulation where the generator costs are randomly varied.   
The sample size used for the Monte Carlo simulation is 50. The percent savings rather than the dollar value are the focus of the paper. 

Two benchmark cases are used to evaluate the savings of the heuristic: Case one considers the initial topology and line constraints whose DCOPF sets an upper bound of the total generation costs, and Case two considers the unconstrained case whose DCOPF provides a lower bound of the generation costs. The cost difference between the two benchmark cases is called the \emph{maximum attainable savings (MAS)}.

The three heuristics are implemented in MATLAB, using  MATPOWER \cite{matpower} as the DCOPF solver. The performance comparison among the three heuristic is detailed in Table 2.

\begin{center}\label{compare}
\begin{tabular}{l@{\hskip 0.05in}c@{\hskip 0.05in}c@{\hskip 0.05in}c}\hline
 heuristic &   saving/MAS  &  lines disconnected & mean effort  \\ \hline 
 Random   &  0.517 $\pm$ 0.092  &   31.22 $\pm$ 6.37  & 0.193\\  \hline
Line Profit    & 0.631 $\pm$ 0.049   &    14.53 $\pm$ 4.12   & 0.317\\ \hline
 Greedy & 0.672 $\pm$ 0.051     &   10.18 $\pm$ 3.98 & 1 \\  \hline
\end{tabular}\end{center}\begin{center}
{Table 2: Performance comparison.}
\end{center}

Strictly speaking, the three heuristics are said to be of the same level computational complexity as they need to redo the   DCOPF for each switchable line  in the worst case. For example, the Random Heuristic or the Line Profit Heuristic may,  though with very low probability,  not be able to find a switchable unprofitable line until after redoing the DCOPF for each allowable line removal.  In general, the Standard Greedy Heuristic, however,   costs much more computational effort than the Random Heuristic and the Line Profit Heuristic especially for a large scale networks. The benefit of the additional computational effort is the increased switching quality, i.e.  higher cost reduction achieved at each stage.  In general, the switching savings in each iteration is the best in Standard Greedy Heuristic, the worst in the Random Heuristic and better in Line Profit Heuristic. Unlike the paradoxical behaviors for specific scenarios, the better switching quality accumulated in each iteration does lead to more average saving for a set of test scenarios as shown in Table 2, which may outweigh the extra computational cost.

It is natural to expect one heuristic may save more if it disconnects more lines.
What seems counterintuitive  is that, on average, the Standard Greedy Heuristic not only attains maximum savings but also disconnects the fewest lines. This, however, can be well explained by the work in \cite{Wang} and \cite{Wang_ar}. 

\begin{theorem}\cite{Wang}
\emph{For an arbitrary current-controlled circuit (a circuit network that is  comprised purely of current sources and resistors),  the total $I^2R$ losses of the network must increase after removing an arbitrary resistance link.}\label{thm1}
\end{theorem}

\begin{theorem}\cite{Wang_ar}
\emph{The increase of total $I^2R$ loss, $\Delta P$, resulting from the  removal of a resistor link with endpoint pair $\{m,n\}$ from a current-controlled circuit is given by $|\Delta P|=|I_{mn}V_{mn}^{'}|$, where $I_{mn}$ denotes the current flowing on the link before its removal, and $V_{mn}^{'}$ denotes the voltage difference between node pair $\{m,n\}$ after its removal.}\label{thm2}
\end{theorem}

Recalling the   equivalence  between  the  current-controlled  circuits  and  DC  power  flow  models  of transmission grids shown in Table 1, Theorem \ref{thm1} tells us that the new transmission  network obtained after line switching  is  of fewer lines but   higher total real $ f^2/B$ (squared line flow over line susceptance) congestion.  Theorem \ref{thm2} quantifies the increase of the $I^2R$ losses of a circuit due to a disconnected resistor   by  using the product of two electrical parameters associated with that disconnected resistor. In the DC  power  flow  models  of transmission grids, it is equivalent to say that the increase of  total real $ f^2/B$ of a transmission grid due to one line removal is given by the product of the power flowing through that  transmission line flow before its removal and the phase difference  between its ending  buses after its removal.

One profitable switching that can greatly reduce the generation cost usually significantly redistributes the power flowing across the network. Such effect is unlikely to be obtained by switching a  transmission line with relatively small pre-removal line flow and post-removal phase difference between its ending buses.  Since the Standard Greedy Heuristic always executes the line removal  leading to maximum savings  in each iteration, it is reasonable to expect that  the average change of  of  total real $ f^2/B$ in each iteration of the Standard Greedy Heuristic  is usually larger than that of the Random Heuristic and Line Profit Heuristic. The combination of the above result together with the ever decreasing total $\overline{f^2}/B$ capacity due to fewer lines in service explain the superior performance of the Standard Greedy Heuristic: not only better approximating the global optimal solution but also better enforcing  network connectivity.

\section{Conclusion}
In spite of the seemingly promising simulation  results of the heuristic approaches pursued by the research community and a certain level of proven predictability \cite{Baillieul,Wang,Wang_ar} of topology control, satisfactory grid-scale solutions are ever-elusive. The cautionary tales associated with our  non-commutativity,  non-monotonicity, and non-consistency paradoxes leads us to wonder whether other paradox -- possibly more problematic -- may associated   with as yet untried heuristics. While optimality for NP-hard problems is not obtained, our simulation results  on a set of iterative heuristics nevertheless offer some justification that, with similar searching space, a greedy heuristic has a better chance in reducing generation cost and enforcing connectivity.   Future work will be focused on further simplifying the problem by developing fast partition algorithms for the decomposition approach \cite{arXiv_vcs}.   Adaptability of the decomposition with respect to various types (traditional \& renewable) of generators will also be the subject of future work.
\bibliography{references}
\bibliographystyle{IEEEtran}
 
\end{document}